\documentstyle[sprocl,psfig]{article}

\def\fr#1#2{{{#1} \over {#2}}}
\def\gsim{\mathrel{\rlap{\lower4pt\hbox{\hskip1pt$\sim$}}
    \raise1pt\hbox{$>$}}}

\def\al{\alpha}

\def\de{\delta}

\def\ve{\varepsilon}

\def\th{\theta}

\def\la{\lambda}

\def\rh{\rho}

\def\si{\sigma}

\def\ta{\tau}

\def\La{\Lambda}

\def\Om{\Omega}

\def\cl{{\mathcal L}}

\def\fr#1#2{{{#1} \over {#2}}}

\def\frac#1#2{{\textstyle{{#1}\over {#2}}}}

\def\lsim{\mathrel{\rlap{\lower4pt\hbox{\hskip1pt$\sim$}}
    \raise1pt\hbox{$<$}}}
\def\gsim{\mathrel{\rlap{\lower4pt\hbox{\hskip1pt$\sim$}}
    \raise1pt\hbox{$>$}}}
\def\sqr#1#2{{\vcenter{\vbox{\hrule height.#2pt
         \hbox{\vrule width.#2pt height#1pt \kern#1pt
         \vrule width.#2pt}
         \hrule height.#2pt}}}}

\def\prt{\partial}
\def\lrpartial{\raise 1pt\hbox{$\stackrel\leftrightarrow\partial$}}

\def\etal{{\it et al.}}

\newcommand{\beq}{\begin{equation}}
\newcommand{\eeq}{\end{equation}}
\newcommand{\bea}{\begin{eqnarray}}
\newcommand{\eea}{\end{eqnarray}}
\newcommand{\rf}[1]{(\ref{#1})}

\def\sech{\mathop{\rm sech}\nolimits}

\begin{document}

\title{Apparent Lorentz violation through spacetime-varying couplings}

\author{RALF LEHNERT}

\address{\vspace{3mm}
CENTRA\\
\'Area Departamental de F\'{\i}sica, Universidade do Algarve, Campus de Gambelas\\
8000-117 Faro, Portugal
\vspace{2mm}\\E-mail: rlehnert@ualg.pt}


\maketitle

\abstracts{
In this talk,
we explore the relation between smoothly varying couplings
and Lorentz violation.
Within the context of a supergravity model,
we present an explicit mechanism
that causes the effective fine-structure parameter
and the effective electromagnetic $\theta$ angle
to acquire related spacetime dependences.
We argue that this leads
to potentially observable Lorentz violation
and discuss some implications
for the Standard-Model Extension.
}

\section{Introduction}

\noindent 
Originally proposed by Dirac,\cite{lnh}
the idea of spacetime-dependent couplings
has remained the subject
of a variety of experimental and theoretical investigations.
As a result of current claims of observational evidence
for a time variation 
of the electromagnetic coupling\cite{webb}
and the recent realization
that such effects are a natural consequence
of many unified theories,\cite{theo}
this idea has regained considerable interest.\cite{jp}
Although there has been substantial theoretical progress 
in the subject,
a realistic underlying theory
allowing concrete predictions
is presently still lacking.
For the identification of high-precision tests,
it is therefore important
to determine generic physical effects
caused by spacetime-dependent couplings.

In models with varying couplings,
the usual spacetime symmetries 
described by the Poincar\'e group can be effected.
For example, 
translational invariance is normally lost in such models.
In this talk, we argue that varying couplings
can also lead to violations
of the remaining spacetime symmetries
associated with the Lorentz group,
a point not widely appreciated.
Intuitively,
this can be understood
when the effective vacuum is interpreted 
as a spacetime-varying medium,
in which,
for example,
isotropy can be lost,
so that
certain rotations,
which are contained in the Lorentz group,
may no longer be associated with a symmetry transformation.

A partial motivation for this study
is provided 
by the extreme sensitivity of experimental Lorentz tests
and by recent progress in the understanding
of a general Lorentz- and CPT-violating 
extension of the Standard Model,\cite{ck}
a framework 
that includes all possible coordinate-invariant 
Lorentz- and CPT-breaking interactions.
It describes the low-energy limit
of possible Lorentz and CPT violation 
at a more fundamental level,
such as strings,\cite{kps} 
nontrivial spacetime topology,\cite{klink}
and realistic noncommutative field theories.\cite{ncqed}
The Standard-Model Extension (SME) 
has provided the basis
for numerous experimental investigations
involving hadrons,\cite{hadronexpt,hadronth}
protons and neutrons,\cite{pn}
electrons,\cite{eexpt,eexpt2}
pho{\-}tons,\cite{cfj,kmm,lipa,edtheo}
muons,\cite{muons} 
and cosmic-ray physics.\cite{rl02}
These studies place tight constraints 
on possible violations of Lorentz and CPT symmetry.
We also remark
that in this context,
the inverse line of reasoning has already been discussed:
certain constant parameters in the SME 
are equivalent to spacetime-dependent masses.\cite{ck}

As part of our analysis,
we construct a classical cosmological solution
in the framework of the pure $N=4$ supergravity 
in four spacetime dimensions
demonstrating how the fine-structure parameter $\al$
and the electromagnetic $\th$ angle
can acquire related spacetime dependences.
Although this model is known to be unrealistic in detail,
it is a limit of the $N=1$ supergravity in 11 dimensions,
which is contained in M-theory.
It could therefore yield valuable insight
into generic features of a promising candidate fundamental theory.
Moreover,
a smoothly varying $\th$ angle
can be associated 
with a Lorentz-breaking Chern-Simons-type interaction.
Our explicit mechanism for a varying $\th$
in the context of a consistent supergravity model
therefore sheds some light
on the usual theoretical difficulties
associated with this term and how they may be avoided.

\section{Cosmology}

\noindent
When only one graviphoton,
$F_{\mu\nu}$,
is excited
and Planck units are adopted,
the bosonic lagrangian 
for the pure $N=4$ supergravity in four dimensions is\cite{cj}
\beq
\cl=\sqrt{g}\left(-\frac 1 2 R
-\frac 1 4 M F_{\mu\nu} F^{\mu\nu}
-\frac 1 4 N F_{\mu\nu} \tilde{F}^{\mu\nu}
+\fr{{\prt_\mu A\prt^\mu A + \prt_\mu B\prt^\mu B}}{4B^2}\right),
\label{lag2}
\eeq
where $g_{\mu\nu}$ represents the metric, 
$\tilde{F}^{\mu\nu}=\ve^{\mu\nu\rh\si}F_{\rh\si}/2$, and
\beq
M=\fr
{B (A^2 + B^2 + 1)}
{(1+A^2 + B^2)^2 - 4 A^2},\quad
N=\fr
{A (A^2 + B^2 - 1)}
{(1+A^2 + B^2)^2 - 4 A^2}.
\label{N}
\eeq
The conventional complex scalar denoted by  $Z$, 
which contains an axion and a dilaton,\cite{cj}
is related to $A$ and $B$
via a canonical transformation,\cite{sugra}
such that
$B$ can be identified with the string-theory dilaton.

Within this model,
one can determine a simple classical solution.
To this end,
we set $F_{\mu\nu}$ to zero for the moment
and assume
a flat ($k=0$), homogeneous and isotropic Universe.
In comoving coordinates,
the associated metric
has the usual Friedmann-Robertson-Walker (FRW) line element
\beq
ds^2 = dt^2 - a^2(t) (dx^2 + dy^2 + dz^2) ,
\label{frw}
\eeq
where $a(t)$ denotes the cosmological scale factor.
The above assumptions also imply 
that $A$ and $B$ can depend on $t$ only.
For a more realistic situation,
the known matter content of the Universe needs to be modeled.
An often employed approach
is to include the energy-momentum tensor of dust, $T_{\mu\nu}$.
If $u^\mu$ is the unit timelike vector
orthogonal to the spatial surfaces
and $\rh(t)$ is the energy density of the dust,
it follows that $T_{\mu\nu} = \rh u_\mu u_\nu$,
as usual.
In the present context,
this type of matter
arises from the fermionic sector
of our supergravity model.
Since the scalars $A$ and $B$
do not couple to the fermion kinetic terms,\cite{cj}
we will take $T_{\mu\nu}$ as conserved separately. 

In this model,
the dependences of $A$ and $B$ on a parameter time defined by 
$\ta=\sqrt{3}/4\;{\rm arcoth}(\sqrt{3c_n/4c_1}\:t+1)$
are given by\cite{sugra}
\beq
A = \pm\la \tanh \left(\fr{1}{\ta} - \fr{1}{\ta_0}\right) + A_0,
\quad
B = \la \sech \left(\fr{1}{\ta} - \fr{1}{\ta_0}\right) ,
\label{be}
\eeq
where 
$\sqrt{3c_n/4c_1}$, $\la$, $1/\ta_0$, and $A_0$ are integration constants.
It can be verified that 
both $A$ and $B$ tend to constant values as $t\rightarrow\infty$.
It follows
that in our supergravity cosmology
the axion $A$ and the dilaton $B$ become fixed 
despite the absence of a dilaton potential.
This essentially results 
from the conservation of energy.

\section{Spacetime-varying couplings}

\noindent
We proceed by considering localized excitations of $F_{\mu\nu}$
in the scalar background given by \rf{be}.
Since experimental investigations
are often performed 
in spacetime regions small on a cosmological scale,
it is appropriate 
to work in local inertial frames.

In the presence of a nontrivial $\th$-angle,
the conventional electrodynamics lagrangian 
in inertial coordinates can be taken as
\beq
\cl_{\rm em} = 
-\fr{1}{4 e^2} F_{\mu\nu}F^{\mu\nu}
- \fr{\th}{16\pi^2} F_{\mu\nu} \tilde{F}^{\mu\nu}, 
\label{em}
\eeq
where $e$ is the electromagnetic coupling.
Comparison with our supergravity model shows that
$e^2 \equiv 1/M$ and $\th \equiv 4\pi^2 N$.
Note that $M$ and $N$
are functions of the axion-dilaton background \rf{be},
so that $e$ and $\th$ acquire related spacetime dependences
in an arbitrary local inertial frame.

The spacetime dependence of both $\al = e^2/4\pi $ and $\th$ 
can be relatively complicated and can vary qualitatively 
with the choice of model parameters.
Figure \rf{fig1} depicts relative variations of $\al$ 
in comoving coordinates 
for $1/\ta_0=0$.  
The fractional look-back time to the big bang 
is defined by $1-t/t_n$, 
where $t_n$ denotes the present age of the Universe. 
The solid line corresponds to no time variation.
Each broken line is associated
with a set of nontrivial choices for 
$\la$, $\sqrt{3c_n/4c_1}\:t_n$, and $A_0$.
Parameter sets 
consistent with the Oklo constraints\cite{oklo} 
are marked with an asterisk. 
Note the qualitative differences in the various plots, 
the nonlinear features, 
and the sign change for $\dot \al$ in the two cases with positive $A_0$. 
Also shown in Fig.\ \rf{fig1}
are the recent experimental results\cite{webb}
obtained from measurements of high-redshift spectra
over periods of approximately $0.6t_n$ to $0.8t_n$
assuming $H_0=65$ km/s/Mpc and $(\Om_m ,\Om_\La)=(0.3,0.7)$.

\begin{figure}
\centerline{\psfig{figure=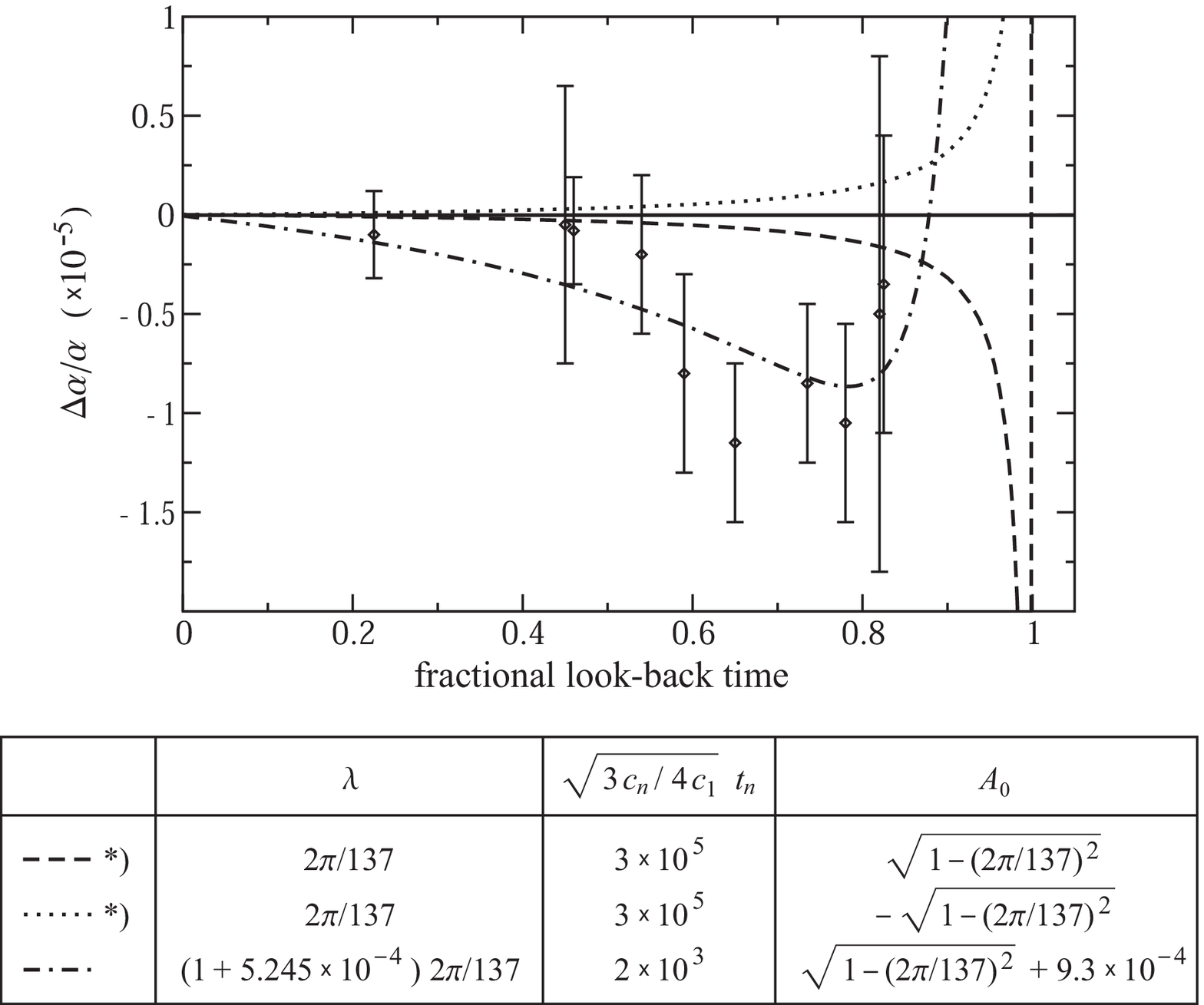,width=0.8\hsize}}
\smallskip
\caption{
Sample relative variations of $\al$
versus fractional look-back time $1-t/t_n$.
}
\label{fig1}
\end{figure}

\section{Lorentz violation}

\noindent
In the presence of charged matter
described by a 4-current $j^{\nu}$,
the equations of motion for $F_{\mu\nu}$ become 
\beq
\fr{1}{e^2}\partial_{\mu}F^{\mu\nu}
-\fr{2}{e^3}(\partial_{\mu}e)F^{\mu\nu}
+\fr{1}{4\pi^2}(\partial_{\mu}\th)\tilde{F}^{\mu\nu}=j^{\nu} .
\label{Feom}
\eeq
Note that in the limit of spacetime-independent $e$ and $\th$,
the usual inhomogeneous Maxwell equations are recovered.
In the present context, however,
the last two terms on the left-hand side of Eq.\ \rf{Feom}
lead to apparent Lorentz violation
despite being coordinate invariant.
This fact becomes particular transparent
on small cosmological scales,
where $\prt_\mu M$ and $\prt_\mu N$
are approximately constant,
and hence,
select a direction 
in the local inertial frame.
As a consequence,
particle Lorentz covariance\cite{ck}
is violated.
It is important to note
that this is not a feature
of the particular coordinate system chosen.
Once $\prt_\mu M$, for example,
is nonzero in one local inertial frame
associated with a small spacetime region,
it is nonzero in {\it all} local inertial frames
associated with that region. 

By contrast,
such Lorentz-violating effects are absent
in conventional FRW cosmologies
that fail to generate spacetime-dependent scalars. 
Although global Lorentz symmetry is usually broken,
local Lorentz-symmetric inertial frames always exist.
It is also important to note
that the above mechanism 
for generating Lorentz-breaking effects 
is not a unique feature of our supergravity model.
Equation \rf{Feom} shows
that any similarly implemented smooth spacetime dependence 
of $e$ and $\th$
on cosmological scales can lead to such effects.
This suggests that this type of apparent  Lorentz breaking
could be a common feature of models
incorporating spacetime-varying couplings. 
We remark in passing 
that the present type of Lorentz violation 
differs conceptually from attempts 
to {\it deform} Lorentz symmetry.\cite{dsr}

An equivalent form 
of the electrodynamics lagrangian \rf{em}
in a local inertial frame
can be obtained via an integration by parts:
\beq
\cl_{\rm em}^{\prime} = -\fr{1}{4 e^2}  F_{\mu\nu} F^{\mu\nu}
+\fr{1}{8\pi^2}(\prt_\mu\th) A_\nu \tilde{F}^{\mu\nu} .
\label{prlagr}
\eeq
The the second term 
on right-hand side of Eq.\ \rf{prlagr}
gives a Chern-Simons-type contribution to the action.
Such gradient-photon couplings 
have been studied previously 
in various contexts.\cite{axph,cptb}
A Chern-Simons-type term 
is also contained in the SME,
and one can identify 
$(k_{AF})_\mu \equiv e^2 \prt_\mu\th/8\pi^2$.
The presence of a nonzero $(k_{AF})_\mu$ 
in \rf{prlagr} shows explicitly 
Lorentz and CPT violation at the lagrangian level.
The situation 
in which $e$ and $(k_{AF})_\mu$ are constant 
has been discussed extensively 
in the literature.\cite{cfj,ck,jkk} 
In this limit,
lagrangian \rf{prlagr} 
becomes translationally invariant.
However, the associated conserved energy 
fails to be positive definite,
which usually leads to instabilities in the theory.
The question arises
how this problem is avoided in the present context
of a positive-definite supergravity model.\cite{fn3}

Although in most models 
a Chern-Simons-type term is assumed to arise
in an underlying framework,
its treatment at low energies
usually involves a constant nondynamical $(k_{AF})_\mu$.
In the present context,
however,
$(k_{AF})_\mu$ is associated
with the dynamical degrees of freedom $A$ and $B$. 
Excitations of $F_{\mu\nu}$
therefore lead to perturbations
$\de A$ and $\de B$ 
in the axion-dilaton background \rf{be}.
As a result,
the energy-momentum tensor $(T^{\rm b})^{\mu\nu}$ 
of the background
receives an additional contribution,
$(T^{\rm b})^{\mu\nu}\rightarrow
(T^{\rm b})^{\mu\nu}+\de (T^{\rm b})^{\mu\nu}$.
It can be demonstrated\cite{sugra} 
that this contribution
does indeed compensate 
the negative-energy ones
associated with a nonzero $(k_{AF})_\mu$.

\section{Summary}

\noindent 
Our analysis suggests
that couplings varying on cosmological scales
can be obtained as simple solutions
of theories beyond the Standard Model
and that such couplings may generically lead 
to  local particle Lorentz violation.
As an illustration,
we have constructed a classical cosmological solution 
within the pure $N=4$ supergravity in four dimensions
that exhibits 
spacetime-varying electromagnetic couplings $\al$ and $\th$ 
and establishes the resulting Lorentz and CPT breaking.
In this model,
a Chern-Simons-type term is generated
but the usual associated stability difficulties
are circumvented.



\begin{thebibliography}{99}
\bibitem{fn3}
A constant timelike $(k_{AF})_\mu$ also violates microcausality.\cite{jkk,caus}
The supergravity cosmology may avoid this,
but a complete analysis of this lies outside our present scope.
\end{thebibliography}
\end{document}